\documentstyle[amsfonts,epsf,prb,aps,multicol]{revtex}
\begin{document}
\draft
\title{Quasi-harmonic vs. ``exact'' surface free energies of Al: a systematic 
study
employing a new interatomic potential} 
\author{U. Hansen,$^{1}$ P. Vogl,$^{1}$ and Vincenzo Fiorentini$^{1,2}$}
\address{{\it 1)\,} Physik-Department and Walter Schottky Institut,
Technische  Universit\"{a}t  M\"{u}nchen, D-85748 Garching, Germany\\
{\it 2)\,} Istituto Nazionale per la Fisica della Materia
and Dipartimento di Fisica, Universit\`a di Cagliari, Italy\\}
\date{to appear on PRB}
\maketitle

\begin{abstract}
We discuss  a computationally efficient classical many-body potential 
 designed to model the Al-Al interaction in a wide range
of bonding geometries. We show that the potential yields results in
properties in excellent agreement with experiment and {\em
ab initio} results for a number of  bulk and surface properties,
among others for surface and step formation energies, and self-diffusion 
barriers.
As an application,  free energy
 calculations are performed for the Al (100) surface by Monte Carlo 
thermodynamic 
integration and the quasi-harmonic approximation. Comparison of the
latter approximation with the reference Monte Carlo results 
provides informations on its range of applicability to surface problems
at high temperatures.
\end{abstract}

\begin{multicols}{2}

\section{Introduction}

Atomistic simulations are playing an increasingly prominent role in
materials science. From studies of crystallization of
clusters\cite{valkealahti} to large-scale simulations of
fracture\cite{zhou} and grain boundary diffusion\cite{harris},
atomistic simulations offer a microscopic physical view  that
cannot be obtained from experiment. Predictions resulting from this
atomic level understanding  are proving increasingly accurate and
useful.\cite{voter} 

 The effective interatomic interaction potential is
the key ingredient in all atomistic simulation. The accuracy of the potential
affects drastically the quality of the simulation result, and its functional
complexity determines the amount of computer time required.\cite{gunsteren} 
Much research effort has therefore
been devoted to the design of potential energy functions.\cite{ping}  
This is especially important in classical dynamics  which, although quantum 
mechanical simulations have been progressing at a rapid pace in recent years,
remains the most (sometimes, the only) affordable way to perform very 
large scale simulations  in materials science. In this paper, we present 
a new carefully designed Al-Al interaction model,  test its performance, 
and  we apply it to the study of  free energies in atomic scale simulations. 

 The ability to compute free energies is essential to understand or
predict many physical phenomena, from the stability of crystal
structures, to the propensity to form defects or disorder, and to
morphology  changes and phase transitions. However, the determination
of free energies from atomic scale computer simulations is a daunting
 task.  
Approximate methods, mostly based on the harmonic vibrational properties
of the system, are commonly in use to this end. Here, we compare
several possible versions of the so called quasi-harmonic
approximation, using as reference  accurate 
 simulation  using canonical or constant-pressure Monte Carlo
methods and thermodynamic integration, focusing on the specific case of
surface free energies. The goal
is to provide a measure of the range of applicability of approximate
methods for complex systems  using a reliable Al interaction model. 

\section{Al Interaction Potential}
\label{potential}

Previously developed interatomic
potentials\cite{cai,hoagland,baskes,mei} for the Al-Al interaction
have mostly focused on bulk and molecular
properties. In this work, we analyze and generalize one of those models with
special regard to surface properties, aiming as usual at describing
 Al in as wide a range of chemical environments as possible, {\it i.e.}
ranging from bulk Al to Al surfaces and surface steps, and to small Al 
molecules. The functionalities of the
refined potential are found to extend significantly those of previous 
ones. 
As we are going to use an embedded atom interaction model, in this Section
 we briefly review the basic ideas of this approach, 
describe the details of the model, and finally assess its quality.

\subsection{Theory}
\label{theory}

In the embedded atom method, each atom in a
solid is   viewed as an impurity embedded in a host comprising
all the other atoms.\cite{daw,daw1} The energy of the host with
impurity is, according to Stott and Zaremba,\cite{stott} a functional
of the unperturbed host electron density, and a function of the
impurity type and position,
\begin{eqnarray}
E = {\cal{F}}_{Z,R}[\rho_h({\bf r})],
\end{eqnarray}
where $\rho_h({\bf r})$ is the unperturbed host electron density, and
$Z$ and $R$ are the type and position of the impurity. Here the energy of
an impurity is determined by the electron density of the host before
the impurity is added. The functional $\cal{F}$ is a universal
function, independent of the host, but its form is unknown.\cite{daw} 
A simple approximation to $\cal{F}$ is the so called local
approximation, whereby the impurity experiences a locally uniform
electron density.\cite{daw1} This can be viewed as the lowest-order term
of an expansion 
involving the successive gradients of the density. The functional
$\cal{F}$ is then approximated by
\begin{eqnarray}
E = F_{i}(\rho _{i}(R_i)) + \frac{1}{2} \sum_j \phi_{ij}(R_{ij}),
\end{eqnarray}
where $\phi_{ij}$ is a pair potential representing the electrostatic
interaction, $R_{ij}$ is the distance between atoms $i$ and $j$, and $F_i$
denotes the embedding energy. The total energy of the system is a sum
over all individual contributions: 
\begin{eqnarray}
\label{e_pot}
E_{\rm tot} = \sum_i F_{i}(\rho _{h,i}) + \frac{1}{2} \sum_{ {i,j} \atop
{i \ne j} } \phi_{ij}(R_{ij}). 
\end{eqnarray} 
A further simplification is introduced assuming that the host
density $\rho_{h,i}$ at atom $i$
is closely approximated by a sum of the atomic
densities $\rho_j$ of the constituent atoms, {\it i.e.}\  $\rho_{h,i} =
\sum_{j,(j \ne i)} \rho_j (R_{ij})$, with $\rho_j$ being the contribution
to the density at atom $i$ from atom $j$. Equation (\ref{e_pot}) is the form
commonly used for molecular dynamics simulations of metals, and is
known as embedded atom potential.

\subsection{Details of the Al-Al interaction potential}

The Ercolessi-Adams interaction 
model for Al was  constructed with the so called force matching
 method\cite{ercolessi}
and, in contrast to most   other empirical models, it gives excellent 
structural and 
elastic  properties for the bulk along with the correct surface interlayer 
relaxations at low-index surfaces.
Furthermore,  we found that the diffusion barriers for surface adatoms
obtained by the Ercolessi-Adams model are in fair agreement with {\em
ab initio} calculations,\cite{stumpf,stumpf1,stumpf2} whereas
those  predicted by most other embedded atom potentials differ 
drastically\cite{liu}
from {\em ab initio} results.
We therefore started off from the Ercolessi-Adams potential to build our own 
refined Al--Al interaction. Without affecting the elastic properties and the 
surface
relaxation properties, we introduced the following modifications to the
 model: 
\begin{enumerate}
\item An additional term was introduced in the pair potential
$\phi_{ij}$ in order to account for an exponential Born-Mayer--like
 repulsion at short Al-Al separation.\cite{abrahamson} This is a key
requirement for studies of
{\it e.g.} physical vapor deposition processes, where the energy of each
single atom easily exceeds the thermal energy by as much as three orders of
magnitude. 
\item In the low density region, three parameters of the embedding
function $F$ were changed in order to improve  several
reference quantities, namely the
Al$_2$ binding energy, and vibrational frequency, and
 the adatom diffusion barrier height on the
Al(111) surface ({\it i.e.} the energy difference between the surface
binding site and the saddle point). 
\item A fifth-order polynomial cut-off function was introduced,
smoothly bringing the potential to zero at an interatomic distance of
5.56 \AA\, (slightly larger than the third-nearest neighbor distance in
bulk Al). 
\end{enumerate}

The total energy  $E_{\rm tot}$ of a system containing Al atoms in an
arbitrary arrangement is written (see Sec.\ref{theory} ) as  
\begin{eqnarray}
\label{e_tot}
E_{\rm tot} = \sum_i F(\rho_i) + \frac{1}{2} \sum_{ {i,j} \atop {i \ne j}
} \overline{\phi}_{ij}(r_{ij}). 
\end{eqnarray}
The atomic density $\rho_i$ in arbitrary units is given as
\begin{eqnarray}
\rho_i = \sum_{j,(j \ne i)} \rho(r_{ij}) \times f_c(r_{ij},R_0,D_0).
\end{eqnarray}
The sum runs over all atoms that lie within the potential range $R_0 + D_0$
(5.56 \AA), which is enforced by the cutoff function
$f_c(r,R,D)$. This function is zero for $r$ exceeding $R+D$ and unity
for $r$ less than $R-D$. For $r$ within the interval $(R-D,R+D)$ it is
defined according to  
\begin{eqnarray}
f_c(r,R,D) =&& -3 \big[ \frac{r-R}{D} +1 \big]^5 + \frac{15}{2}
\big[\frac{r-R}{D} +1 \big]^4  \nonumber \\ &&- 5 \big[ \frac{r-R}{D}+
							1 \big]^3 + 1.
\end{eqnarray}
The function $\rho(r)$ in Eq.(\ref{e_tot}) is spline-interpolated
using the values reported in Table \ref{rh}; the parameters $R_0$ and $D_0$
are given in Table \ref{parameters}. The embedding function $F(\rho)$
is also spline-interpolated, and the corresponding values for $F(\rho)$
are collected in Table \ref{uu}. 

The pair potential term in Eq.(\ref{e_tot}) is written according to
\begin{eqnarray}
\overline{\phi}_{ij} = \big[\; \phi(r_{ij}) + (A\; \exp\{-\lambda
r_{ij}\} \times  f_c(r_{ij},R_\phi,D_\phi) \nonumber \\  
-B)\; \big] \times f_c(r_{ij},R_0,D_0). 
\end{eqnarray}
The function $\phi$ is tabulated in Table \ref{v2}.
The first cut-off $f_c(r_{ij},R_\phi,D_\phi)$ switches on the exponential
repulsive term at small distances ($r <$ 2.25 \AA), while
$f_c(r_{ij},R_0,D_0)$ terminates the interaction range of the
potential. The corresponding parameters are given in Table
\ref{parameters}. The exponential term ensures that one gets
a Born-Mayer repulsion at  short separations for, {\it e.g.},
diatomic molecules.\cite{abrahamson}

\subsection{Assessment of the potential}
\label{assesment}

We now present evidence that the potential just described
 yields satisfactory results for a variety of properties of
Al in different
environments. In particular we address the bulk, the dimer,
low-index surfaces and steps thereon, and self-diffusion
on different low-index surfaces;  also included in this test section
is  the energy dependent sticking coefficient of high energy 
Al atoms on Al (111). The changes to the potential significantly improved
agreement with experiment and other theoretical predictions in several 
instances where errors were typically of order 50\%.

\subsubsection{Bulk properties}

By construction our model does not alter the equilibrium lattice
constant $a_0$, the cohesive energy E$_{\rm coh}$  and the elastic
properties of the previous model of Ercolessi. We obtain as in
Ref. \onlinecite{ercolessi} $a_0$ = 4.03 \AA, E$_{\rm coh}$ = 3.36 eV, C$_{11}$
= 118 GPa, C$_{12}$ = 62 GPa and  C$_{44}$ = 36 GPa. For comparison,
the experimental values\cite{kamm} are  C$_{11}$ = 114 GPa , C$_{12}$ = 62 GPa
and  C$_{44}$ = 32 GPa,  and LDA calculations\cite{stumpf2,deyirmenjian}
predict
 $a_0$ = 3.98 \AA, E$_{\rm coh}$ = 4.15 eV, C$_{11}$ = 135 GPa, C$_{12}$ = 70 
GPa
and  C$_{44}$ = 35 GPa.

\subsubsection{Diffusion barriers}

Diffusion is central to many physical processes which determine the
morphology of surfaces, such as step flow, nucleation, and
growth.\cite{boisvert} 
It is of obvious importance to study diffusion processes
theoretically, since 
direct observations of surface diffusion by
means of  field ion microscopy (FIM)\cite{kellog} are limited to a few
surfaces due to  the response limits of the materials of interest to high
voltages.\cite{boisvert}   The barriers for
single adatom diffusion on Al surfaces
 calculated by Stumpf and Scheffler\cite{stumpf,stumpf1,stumpf2} using
{\em ab initio} LDA techniques, provide a stringent test for the
present empirical Al-Al interaction model.  It is generally accepted
that
diffusion on flat metal surfaces proceeds by either hopping
or exchange.\cite{kellog1} In the two following subsections
compare the results of the present model for these  mechanisms with 
previous LDA results. 

{\em Hopping Diffusion --} 
During hopping diffusion the adatom is moving between minima of the
potential energy surface, {\it i.e.}\ between stable or metastable binding
sites. On the (111) surface the stable adsorption sites are the 3-fold
fcc and hcp sites; on the (100) surface there is single independent
adsorption site, the four-fold
hollow;  the (110) surface is analogous, with  a five-fold
site. Hopping diffusion on 
the (110) surface is intrinsically anisotropic, since it can proceed
 perpendicular or parallel to the
[1$\overline{1}$0]-oriented atomic rows,  respectively 
via the short bridge or long bridge paths. The activations energies for
the long and short bridge  are labelled $E_\|$ and $E_\bot$
respectively. 

For each surface we performed total energy calculations for the adatom
sitting at the adsorption site and at the bridge site. At the latter 
site the total energy is minimized with respect to distance of the
adatom from surface. All other Al positions are fully optimized. The 
energy difference 
between adsorption  and bridge site is defined to be the
activation energy for hopping diffusion. Further technicalities are discussed 
in
the Appendix.
Table \ref{diffusion_barriers} summarizes the activation energies for
surface self diffusion obtained with the present model and compares
the barrier heights to results from previous {\em ab initio}
calculations.\label{stumpf2,freibelmann} The barrier for diffusion
on Al (111) was used to set up the potential, hence it agrees
 with  LDA results by construction.  On Al (100),
 we obtain a diffusion barrier  in good agreement
with first principle calculations; we are not aware of 
experimental results for these self-diffusion barriers
 on Al(111) and Al(100).
 According to a recent study,\cite{feib3}
the hopping-self-diffusion barriers calculated ab initio in the
 generalized gradient approximation to density functional theory
  for unreconstructed fcc(100) surfaces equal
one sixth the bulk cohesive energy; this is found to be the case 
for Al also in our calculations.
In the case of Al (110),
the present potential correctly predicts diffusion anisotropy, although
 some quantitative discrepancy exists with the predictions of {\em
ab initio} calculations.\cite{stumpf,stumpf1}
FIM studies\cite{kellog}
 find no anisotropy for the diffusion on the (110) surface, and
 report a barrier height of 0.43 eV  for both paths.
This discrepancy with our result is due to the fact that
here we only addressed hopping diffusion
for demonstrative purposes. 
 In fact, exchange diffusion normal to the rows is known to have a
 barrier as low as $\simeq$ 0.6 eV,  \cite{stumpf,stumpf1}
 which  restores a reasonable agreement with experiment.

{\em Exchange Diffusion --}
 Diffusion by atomic exchange occurs as  the adatom replaces a surface
atom, which in turn pops up at an adjacent stable surface
 site. Diffusion by exchange was discussed by Bassett and
 Webber\cite{bassett} and Wrigley and Ehrlich\cite{wrigley},
and  predicted theoretically for (100) surfaces 
 by Feibelman.\cite{feibelman} This
 diffusion mode can lower significantly the effective diffusion
 barriers. Views on why diffusion by exchange is favorable for some metal
 surfaces is discussed in Refs. \onlinecite{stumpf2}, 
\onlinecite{feibelman}, \onlinecite{byung}, \onlinecite{feib2},
and references therein.
 For  Al (100), we calculated the activation energy for exchange diffusion
 (see Table \ref{diffusion_barriers}) and found it lower than for hopping,
as do first-principles calculations. The quantitative agreement is 
reasonably good.

\subsubsection{Surface and step energies}
\label{surface_formation_energies}

The surface energy, defined as the difference between the energy of
an atom at the surface and in the bulk environment, is usually calculated as
\begin{eqnarray}
E_{\rm surf}^{\rm area} = \frac{E_{\rm slab} - N \;E_{\rm bulk}}{2 A} \\
E_{\rm surf}^{\rm atom} = \frac{E_{\rm slab} - N \;E_{\rm bulk}}{2 N_{\rm 
surf}}
\label{surf_formula} 
\end{eqnarray}
where $E_{\rm slab}$ is the total energy of the slab, $N$ the total number
of atoms in the slab, $N_{\rm surf}$ the number of atoms on each surface,
$E_{\rm bulk}$ is the total energy per bulk atom, $A$ is the
area of each free surface of the slab, and the factor 1/2 accounts for
the two free surfaces of the simulation cell;  periodic boundary
conditions are applied in the planar directions. These formulas are
not problem-free in general,\cite{f-m} but we have checked  that they
are in the cases of interest to us.

 We calculated the
formation energies for the low index (111), (100) and (110) Al
surfaces with the present Al-Al interaction model. The comparison of our
results to those obtained in {\em
ab initio} LDA investigations, given in Table
 \ref{surface_energies} shows that
the trends of surface energies of our Al model are consistent with the
{\em ab initio} calculations. We obtain all surface energies about 20\%
lower than the LDA surface energies. Keeping in mind the known LDA
overestimate of the binding energies (the LDA  cohesive
energy of Al, 4.15 eV,\cite{stumpf1} is about 20\% larger than the
experimental value of 3.36 eV) the agreement between the two approaches
is in fact excellent.

\narrowtext
\begin{figure}[t]
\epsfysize=7.5cm
\centerline{\epsffile{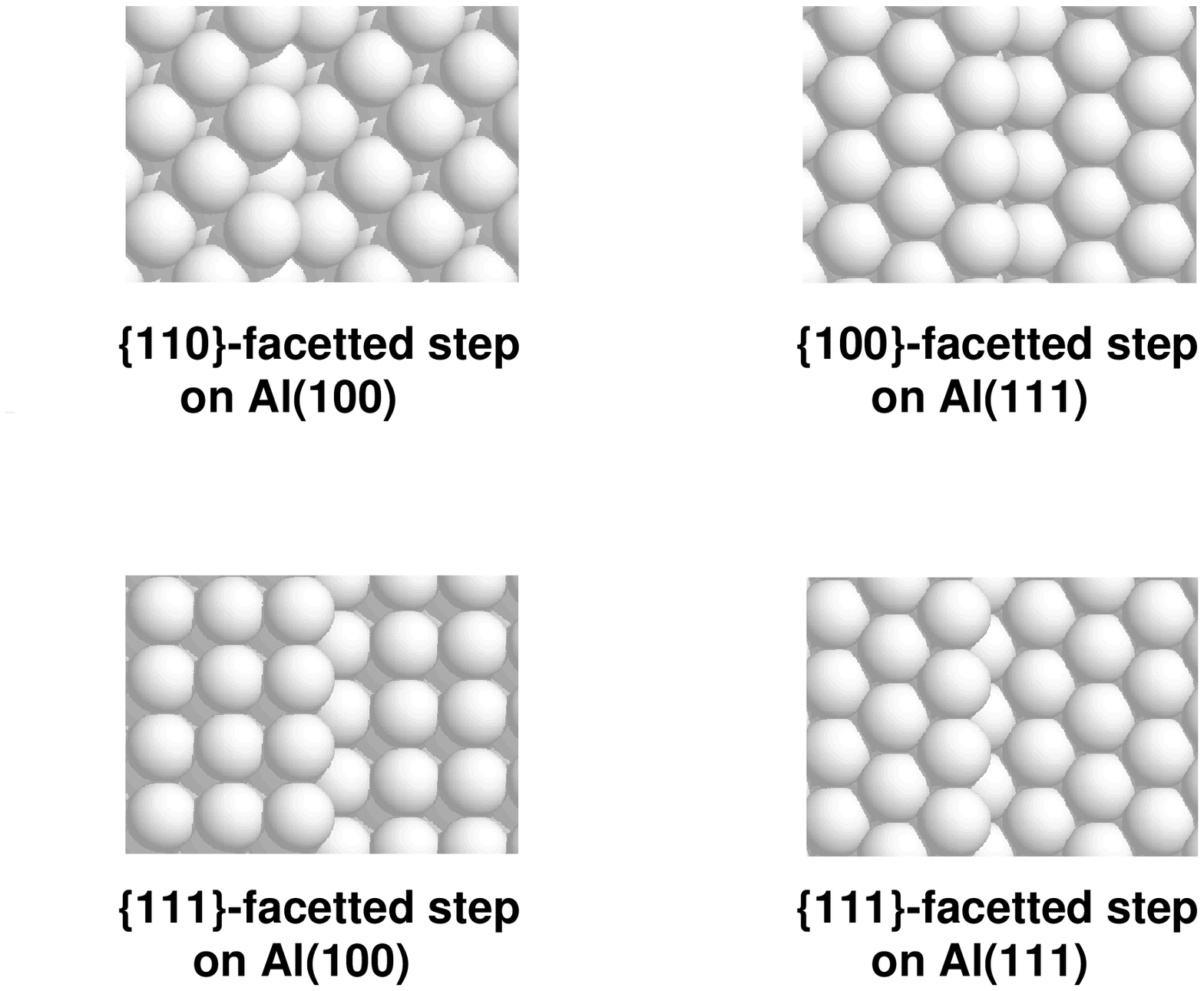}}
\caption{\label{figure_1}
Atomic arrangement of the \{110\} and \{111\} facetted steps on
Al (100), and the \{100\} and \{111\} facetted step on Al (111).} 
\end{figure}

The energy per unit length of a step on a low-index surface
  is defined in terms of the 
energies of the low-index face and the vicinal surface used to 
simulate the step itself, and of the the geometrical parameters
  thereof:\cite{phil} \begin{eqnarray}
E_{\rm step} =  d_{s-s} E_{\rm vicinal} - l_{\rm terrace} E_{\rm low-index},
\end{eqnarray}
with $d_{s-s}$ the step-step distance, and $l_{\rm terrace}$ the
terrace length on the vicinal surface.
In analogy to the surface energy, the step energy can also be 
expressed per step atom.  

Although a stepped vicinal surface can be specified by its
corresponding Miller indices, this notation is not very convenient, as
it does not indicate at first sight the geometrical structure of the
surface. Thus we use instead the notation $[n(h,k,l) \times
m(h^\prime,k^\prime,l^\prime)]$ by Lang {\it et al.},
\cite{lang,desjonqueres} where $(h,k,l)$ and
$(h^\prime,k^\prime,l^\prime)$ are the Miller indices of the terraces
and  ledges respectively; $n$ gives the number of atomic rows in
the terrace parallel to the step, and $m$ corresponds to the height of the
step. In the case of monoatomic-height steps, $m$ is
omitted in this notation.

For the two low index surfaces Al (100) and Al (111),  we calculated
the  formation energy of different steps. On Al (100)
there exist two monoatomic steps, the close-packed $\{111\}$-facetted and the  
more open $\{110\}$-facetted. The former belongs to the family of
$(1,1,2n+1)$-surfaces, the latter to the family of the
$(1,0,n)$-surfaces.\cite{eisner,hove} For the calculation of the step
energies we used the Al$(1,1,15)$ = Al$[9(100) \times (111)]$ and the
Al$(1,0,9)$ = Al$[9(100) \times (110)]$ surfaces. 

On Al(111) there are
two types of close-packed steps, the $\{111\}$-facetted and the
$\{100\}$-facetted. The corresponding vicinal surfaces belong 
to the $(n,n,n-2)$ and $(n,n,n+2)$ families,
 respectively.  We used the Al$(9,9,7)$ =
Al$[9(111) \times (111)]$ and the  Al$(8,8,10)$ = Al$[9(111) \times
(100)]$ surfaces. The geometry of the different steps on the A$l(100)$
and Al (111) surface is depicted in Fig.\ref{figure_1}. For all these
vicinals, the terraces separating the steps have the same width of 9
atomic rows. We have verified that step-step repulsion at these
inter-step distances is already in the long-range elastic regime
$\sim d_{s-s}^{-2}$,
\cite{phil,phil2} the steps are far enough to extract their formation energy
without an unknown bias from the interstep interaction.

 Table \ref{step_energies}
lists the results for step formation energies,
 and compares them to {\em ab initio} data. The empirical Al 
potential describes the step formation energies for the two different steps on
Al (111)  in excellent agreement with first-principles
calculations. More energy  is needed 
to  create  steps on the close-packed Al (111) surface
 than on the more open Al (100). The open step on the
Al (100) surface has a 20\% larger formation energy than the
close-packed step,  in agreement with bond cutting arguments.\cite{meth}

\subsubsection{The Al Dimer}

The dimer  is a stringent test for an Al-Al interaction
model, since atoms in a dimer experience a very different chemical
environment compared to bulk or surface atoms. For Al$_2$,
 our model yields a binding energy per atom of 0.70 eV and
a bond length of 2.70 \AA. 
The binding energy was indeed  used as  input to
determine the model parameters,  matching the LDA value \cite{robertson}
of 0.71 eV  ($-0.68 \pm 0.03$eV experimental, Ref. \onlinecite{fu})  The lowest 
vibrational frequency is calculated to
be $\nu$  = 290 cm$^{-1}$, in excellent  agreement with the
experimental value \cite{djugan}   of 284.2 cm$^{-1}$.
The predicted bond length of our model matches exactly the
experimental estimate.\cite{djugan} Thus, the present model describes
satisfactorily  the bonding of Al in the rather
extreme case of the Al$_2$ dimer. 

As the understanding of diffusion and
growth requires a knowledge of the binding energies of small aggregates of
adatoms, we also calculated the energy of two Al adatoms
sitting at neighboring fcc sites on an Al (111) surface. The energy
gain with respect to isolated adatoms is $0.50$ eV, {\it i.e.}\ the Al
ad-dimer is stabilized appreciably over separated adatoms.
 LDA calculations\cite{stumpf2} yield a similar energy gain
of $0.58$ eV.

\subsubsection{Sticking coefficient for hyperthermal Al atoms}

\begin{figure}[ht]
\epsfysize=5cm
\centerline{\epsffile{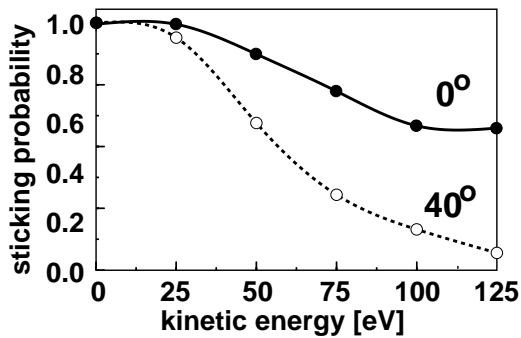}}
\caption{\label{figure_2}
Sticking probabilities for hyperthermal Al atoms impinging on an
Al (111) surface. As a function of kinetic energy the solid line with
filled circles depicts the reaction probabilities for Al atoms
impinging normal on the surface and the dashed line with the open
circles show the latter quantity for an angle of 40$^{\circ}$ to the
normal.} 
\end{figure}

During physical vapor deposition the Al atoms emitted from the sputter
source have a non-thermal energy distribution, with kinetic energies
 exceeding 10 eV.  Therefore the sticking coefficient, a key ingredient
for a reliable modelling of metal film growth,  cannot be assumed to
be constant and independent of the particle's energy as it is
typically done. In order to elucidate the dependence of the
sticking coefficient on impingement energy, we start our simulations with the
incident Al atom placed outside the interaction range of the
surface. Its initial  kinetic energy is set in the range of 0 to 125
eV, and its starting angle off the  surface normal in the range
$0^{\circ}$ to $60^{\circ}$, which corresponds to typical ionized
physical vapor deposition conditions. The trajectories of the incident
atom, and of any other atom which may be etched away from the surface
upon impact, are  monitored until either  a certain time span has
elapsed, or the outcoming atoms (in the case of reflection or etching)
have traveled  a distance of 10 \AA\, away from the surface. Analyzing
200 trajectories per incident energy and angle, we collected a
statistically significant sample of well-defined adsorption,
reflection, and etching  events. The relative probability of the
sticking coefficient is calculated as the ratio of the number of
adsorption events to the total number. The typical  statistical error
in the reaction probability thus determined is below 5 \%.
Fig.\ref{figure_2} depicts
the sticking coefficient as a function of energy for Al atoms
imping normally on the surface (solid circles) or
at an off-normal angle of 40$^{\circ}$ (open circles). 
 The sticking probability 
varies strongly with the incident kinetic energy; the angle to the normal
also has a drastic effect. Details of the
molecular dynamics calculations are given in
Sec.\ref{stickingdetails}. Further discussion and results
 on high energy deposition are reported in Ref. \onlinecite{rapid}.

\section{Application: Free energy calculations}
\label{free}

In this applicative Section of the paper, we compare surface free
energies of Al computed using different levels of quasi-harmonic
approximation,
and thermodynamic integration via Monte Carlo simulations.
The latter effectively functions as ``exact''  reference for the
various harmonic approximation.  Before presenting the results, we 
briefly review the background theory of the different approaches.

\subsection{Theory}

{\em Thermodynamic Integration -- } The free energy cannot be calculated
as an ensemble average.\cite{frenkel} The method of thermodynamic
integration~\cite{mei,frenkel,foiles1,foiles2,morris,vlot} circumvents
this problem starting from the
concept of Stockmayer fluid, a fictitious system in which the 
 interparticle interaction potential $U_{\lambda}$ is gradually switched 
on from a known reference potential $U_h$ to the actual, full interaction
potential $U$; the mixing of $U_h$ and $U$  into the effective potential  
is  controlled by a parameter $\lambda$:
\begin{equation}
U_{\lambda} = (1-\lambda)U_h + \lambda U.
\end{equation}
The key relation of the method concerns
the derivative of $F$ with respect to $\lambda$:
\begin{equation}
\frac{d F}{d \lambda} = \langle \frac{d U_{\lambda}}{d \lambda} \rangle_\lambda 
= \langle U - U_h \rangle_\lambda.
\end{equation}
 The subscript $\lambda$ means that the average has to be evaluated
with the interaction potential $U_{\lambda}$. Integrating the latter equation 
one arrives at the following expression for the free energy of the system
of interest: 
\begin{equation}
F_{\lambda=1} = F_{\lambda=0} + \int_0^1 \langle U -U_h
\rangle_\lambda d\lambda \label{lambda_int} 
\end{equation}
The usual choice for the  reference system is the Einstein crystal 
({\it i.e.} a system of non-interacting harmonic
oscillators with the
interaction potential $U_{h} = \frac{1}{2} m\omega_D\sum_i |{\bf{r}}_i
- {\bf{r}}_{i0}|^2$), whose free energy is   
\begin{equation}
F_{\lambda=0} = -3N k_B T \ln \Big( \frac{T}{\Theta_D} \Big),
\end{equation}
with  $\Theta_D$ the Debye temperature ($394$ K for 
Al~\cite{launay}). The free energy of the real system
 can thus be obtained at any given
temperature by a series of canonical Monte Carlo simulations.

A faster way to obtain the temperature variation of the free energy is to
integrate the thermodynamical relation
\begin{equation}	
\frac{d}{d T} \left( \frac{F}{T} \right) = - \langle \frac{H}{T^2}
\rangle, \label{T_variation} 
\end{equation}
from a reference temperature upwards. This
 requires a simple ({\it e.g.}) Monte Carlo ensemble average
of the energy for each temperature, and of course a reference value of
$F$ from thermodynamic integration.

In summary, the free energy of the system
at a reference temperature $T_{0}$ is determined with 
Eq.(\ref{lambda_int}); then, the temperature variation from $T_0$ to $T$
is calculated using
Eq.(\ref{T_variation}). Both steps were performed   by
 canonical  Metropolis Monte Carlo simulations.
\cite{frenkel,allen} As detailed below, thermal expansion is taken into 
account performing the  NVT Monte Carlo calculations at
 the temperature-dependent
lattice constant determined independently by NPT molecular dynamics.

{\em Quasi-harmonic approach --} A popular approach to  free
energy calculations  is the quasi-harmonic approximation. Thereby, the
full interatomic potential is replaced by its quadratic expansion 
 about the atomic equilibrium positions. The system is
then equivalent to a collection of harmonic
oscillators, and diagonalization of the corresponding dynamical matrix
yields the squares of the normal-mode  frequencies, {\it i.e.} the phonon
spectrum. In bulk systems the
dynamical matrix is a $3\times3$ matrix; for a slab system it is a
$3\ell\times3\ell$ matrix, where $\ell$ is the number of atomic layers in the
slab. The dynamical matrix is given by\cite{allen-dewette}
\begin{equation}
D_{\alpha \beta}({\bf \ell}\; {\bf \ell}^\prime) = \frac{1}{m}
\sum_{\ell^\prime} \Phi_{\alpha \beta}({\bf \ell}\; {\bf \ell}^\prime)
\exp[i{\bf q}({\bf r}_0 - {\bf r}_0^\prime )] 
\end{equation}
where the force constant matrix $\Phi_{\alpha \beta}({\bf \ell}\; {\bf
\ell}^\prime)$ is defined as 
\begin{equation}
\Phi_{\alpha \beta}({\bf \ell}\; {\bf \ell}^\prime) = \left( \frac{
\partial^2 U}{\partial u_\alpha({\bf \ell}) \; \partial u_\beta({\bf
\ell}^\prime)} \right)_0 . 
\end{equation}
The subscript "0" indicates that the second derivatives are to
evaluated at the true mean positions of the atoms, with any
displacements from the bulk positions ({\it e.g.} surface relaxations
or reconstructions)  taken into account. The
equilibrium positions $x_0^\ell,y_0^\ell,z_0^\ell$ of the atoms are given by
the vectors ${\bf \ell}=(\ell_1,\ell_2,\ell_3)$; the $\ell_3$ axis is
perpendicular to the surface and the position of an atom within a
plane is specified by $\ell_1,\ell_2$, and $u_\alpha ({\bf \ell})$ describes 
the
$\alpha$ component ($\alpha = x,y,z$) of the position of the $\ell$-th atom
from its mean position $x_0^\ell,y_0^\ell,z_0^\ell$. 

\begin{figure}
\epsfysize=10cm
\centerline{\epsffile{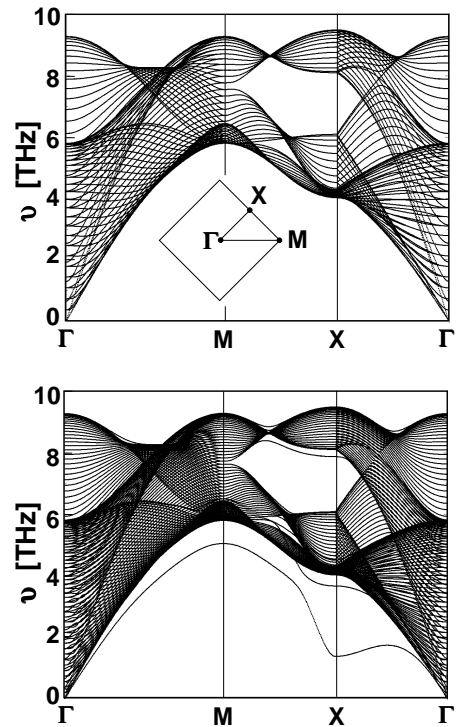}}
\caption{\label{figure_3}
Bulk (a) and surface (b) phonon spectra for Al calculated with a 50
layer slab. In (b) the slab has two  (100) surfaces, the frequencies
are plotted along lines of high symmetry. The corresponding
2d-Brillouin zone is shown in the inset of panel (a).} 
\end{figure}

The phonon spectrum of  bulk Al and of Al (100) are
displayed  in  Fig.\ref{figure_3}, upper and lower panels respectively
(for both calculations supercells comprising 50 layers stacked along (100)
 have been employed).   A variety of
surface modes appear in bulk gaps or split off from  bulk band
edges. These additional modes are the source of the different 
vibrational free energy of surface systems in comparison to
bulk systems. The free energy in this approximation is calculated for
lattice and geometrical parameters  {\bf a} at  temperature {\it T} as
\begin{equation}
F({\bf a},T) = E_0({\bf a}) + k_B T \sum_{{\bf k},j} \ln\,
(2\, \sinh \frac{\hbar
\omega_j({\bf k})}{2 k_B T}).
\label{fat} 
\end{equation}
 The sum runs over all phonon  polarizations $j$ and wave vectors ${\bf k}$ in 
the
Brillouin zone, with $\omega_j({\bf k})$ the frequency of the
corresponding modes. Both the frequencies, and the internal energy
$E_0({\bf a})$ of  the ideal static lattice, depend on
all the lattice and geometrical parameters ${\bf a}$. The latter include
the bulk lattice constant and, for the surface, the additional geometrical
parameters involved in relaxations or reconstructions.

In a bulk system  the forces on each atom are
 zero by symmetry, independently of $a$, so that it is strictly
correct to neglect
 the first derivatives in the quadratic expansion of
the potential energy. We have verified that the quasi-harmonic
 approximation does indeed work very well for the bulk even in
 comparison to thermodynamic integration.
 For a surface, the situation is different, since
 the interlayer spacings (expecially those of the top surface layers)
 will change with respect to the bulk,
{\it i.e.}\ the surface will generally either contract or expand. The 
 average equilibrium positions of the near-surface layers 
 (the interlayer spacings in the case of simple
relaxation)   are determined by the minimum of the free energy, but that 
is, of  course, unknown a priori. In addition, the harmonic expansion
is not strictly correct (since the forces, {\it i.e.} the derivatives of the
internal potential energy are not zero at the free energy minimum), which
is why one calls this the quasi-harmonic approximation to begin with.
In fact, it is clear that there are  several levels of approximation
possible for the quasi-harmonic approach; here we consider some of those:
\begin{enumerate}
\item The computationally simplest way is to optimize
the  atomic configuration and calculate the phonons {\it at zero temperature},
  and evaluate the free energy vs. $T$ using those ingredients for all $T$.
  Within this approach, the
quasi-harmonic approximation is strictly valid, as we expand the
potential energy function around the equilibrium positions, and the $T$
dependence enters solely with Eq.(\ref{fat}). In real systems, of course,
both the  surface internal energy {\it and} the vibrational contribution to the 
free
energy will change with $T$, but it is a priori unclear to what degree
this influences the result.  
\item Another way to account for the effects of finite
temperature is to take the $T=0$ atomic positions, rescale their
coordinates as a function of the temperature
according to the appropriate bulk thermal expansion
coefficient, and recalculate the phonons (and hence the free energy)
 for the expanded lattice.
This is a hybrid case in which $T$ not only affects the force constants, 
but also the surface internal energy. Of course it is arbitrary to
use the scaled $T$=0 interlayer spacings at non-zero temperatures. Also,
it should be kept in mind that the harmonic approximation is
not strictly valid for the expansion of the potential energy around 
non-equilibrium positions.  
\item A further possibility is to
rescale all the coordinates according to thermal expansion first,
 and then re-optimize all atomic positions; the phonons and the free energy
are calculated for that geometry. Here one is consistent
with the prerequisites of the quasi-harmonic expansion, but  at
the cost of getting  wrong interlayer spacings at the surface. 
\item The real thing is of course to minimize the total free
energy with a ``self-consistent'' adjustment of the 
atomic positions of all layers in the slab system, resulting in the
thermodynamic equilibrium configuration of the surface system. In
practice, one starts with the  bulk positions rescaled according to
thermal expansion, and then adjusts the
interlayer spacings of a few near-surface layers to obtain the minimum
of the free energy. The major contribution is generally due to the 
first two surface layers, the only having a sizable displacement from the
bulk interlayer
spacing. In our calculations, we therefore changed 
$d_{12}$ by $\pm 3\%$ and $d_{23}$ by $\pm 2\%$. 
\end{enumerate}

\subsection{Results}
We now compare the free energy
of an Al(100) surface calculated within the different quasi-harmonic
 approaches (1--4) described above,
with the results of thermodynamic
integration; the latter  effectively functions as exact reference
since it takes the full potential into account, hence in particular
all anharmonic contributions. We chose  
the (100) surface for demonstrative purposes, as intermediate between
 the closed packed (111) and the more open (110) surface. 

\subsubsection{Comparison of different methods}
In order to obtain the bulk lattice constant at different temperatures
we first performed zero-pressure molecular dynamics
simulation\cite{allen,berendsen,martyna} using our Al potential.
 In the linear regime the
expansion coefficient is $\alpha = 1.64 \times 10 ^{-5}$
\AA/K,  the experimental value being $2.36
\times 10 ^{-5}$ \AA/K.\cite{pearson}
  Deviations from linearity set in \cite{expansion}
at about 500 K. 
 The dimensions of the simulation
cell with periodic boundary conditions 
correspond, in all subsequent calculations,
 to the bulk lattice constant at the relevant
temperature. The surface free energy calculations imply 
the evaluation of  the bulk free energy, and of  the free 
energy of a slab system with two surfaces. The surface free
energy is then determined with Eq.(\ref{surf_formula}). 

\begin{figure}[ht]
\epsfysize=6cm
\centerline{\epsffile{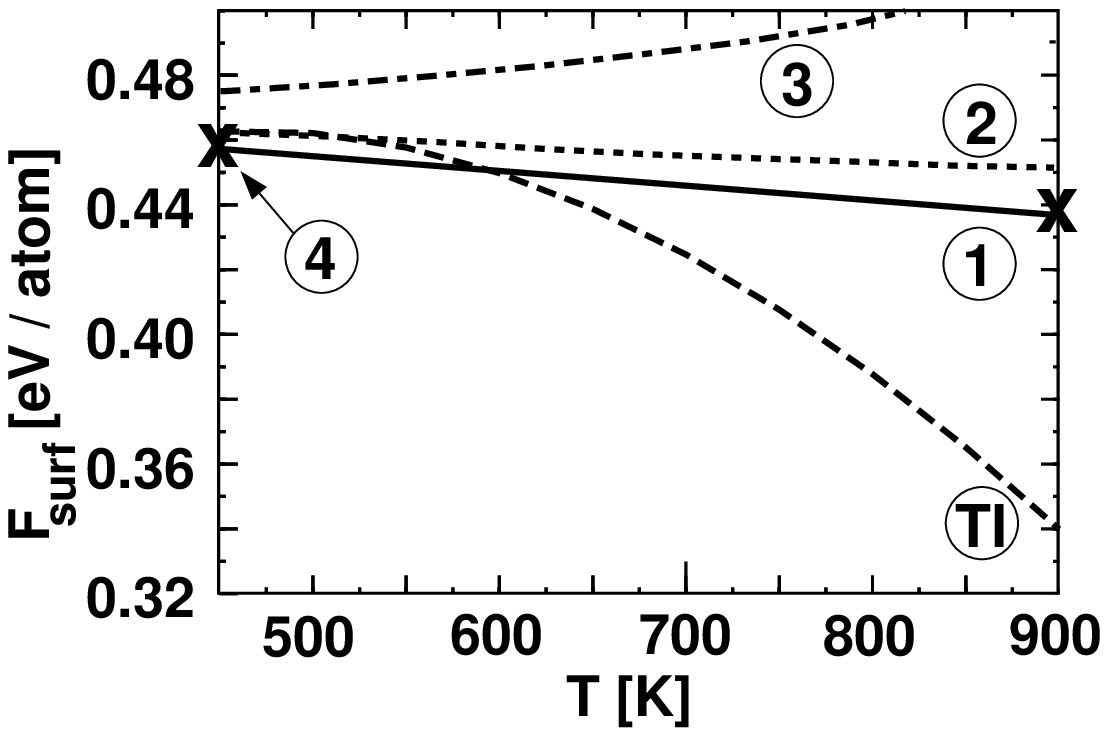}}
\caption{\label{figure_4}
Surface free energies for Al (100) calculated with different
quasi-harmonic approaches (as discussed in the text) and with the method of 
thermodynamic integration. Solid line:
approach (1), zero-temperature phonons for all temperatures;
dotted line: approach (2), positions rescaled according to thermal expansion;
dash-dotted line: approach (3), as (2) with re-optimization of atomic
positions;
crosses: approach (4),  minimization of the free energy in the 
\{$d_{12},d_{23}$\} plane; dashed line: (TI) 
thermodynamic integration, reference for harmonic approximations. See text for 
more details.}
\end{figure}

In Fig.\ref{figure_4} we compare the surface free energy 
calculated with thermodynamic integration and  the quasi-harmonic approach. All
versions of the latter underestimate severely the
 temperature variation of the surface free energy. This
is mainly due to the neglect of anharmonicity, which is also responsible for
thermal expansion. A first important result is then that at
 temperatures $T\geq \Theta_D$,  the harmonic approximation is inappropriate
 for Al surfaces. 

Note that the failure of the
harmonic approximation for   
the present relatively high-temperature calculations does not 
affect the successes of this approach at low temperatures,
an example being the recent first-principles calculations for 
Be surfaces.\cite{degi} The reason why those results are compatible with ours 
is clearly that we work well above the Debye temperature of
our system ($\sim$ 400 K), whereas the highest temperature
considered in Ref. \onlinecite{degi} is 750 K, well below 
$\Theta_D^{\rm Be} \simeq$ 1000 K (as extracted from a Debye-Einstein 
model). Of course, the quasi-harmonic approach will generally fail 
if applied to systems at sufficiently high temperatures.

To sort out the relative merits of the various levels of harmonic 
approximation,
we focus on the effects of the interlayer spacing $d_{12}$ (between 
first and second layer) and $d_{23}$ (between second and third layer) 
on the  surface energy and on the vibrational
contribution to the surface free energy. Essentially this is the fourth level 
of 
approximation  mentioned earlier.  We pick $T$=450 K for
demonstrative purposes, and expand the lattice accordingly.

\begin{figure}
\epsfysize=10cm
\epsfxsize=6cm
\centerline{\epsffile{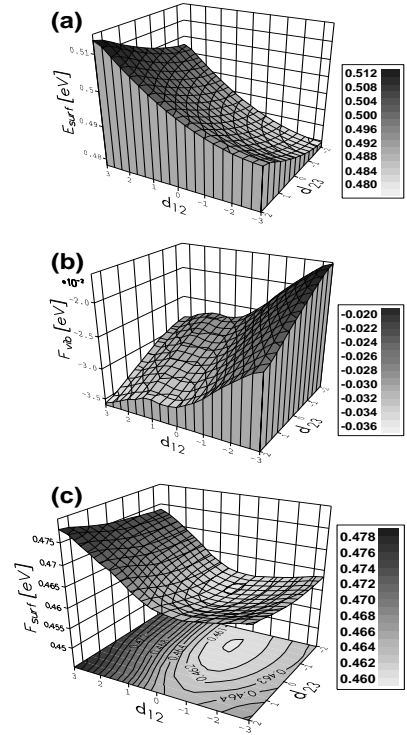}}
\caption{\label{figure_5}
Panel (a): dependence of the plain surface energy on the
interlayer spacing  $d_{12}$ and $d_{23}$. Panel (b):
vibrational contribution to the surface free energy as a function of
the interlayer spacings. Panel (c): total surface free energy, sum of the
two previous contributions.}
\end{figure}

 Panel (a) in Fig.\ref{figure_5} shows
the variation of the plain surface energy as a function of the interlayer
spacings (expressed in turn in percentage of the bulk interlayer spacing).
An increase of the interlayer spacings tends to increase the energy 
drastically,
a decrease to reduce it. The  minimum at  around --3\% for both spacings.
These fairly unrealistic values result from the optimization of the interlayer 
spacing for a laterally expanded surface.

The excess surface free energy, {\it i.e.} the vibrational contribution 
is shown in panel (b) of Fig.\ref{figure_5}. An increase of the interlayer
 spacings leads to softer force constants and hence to lower frequencies,
 which yield according to Eq.(\ref{fat})  a more negative value of the
surface excess free energy. The dependence on the first spacing is stronger 
than on the second, although conceptually both spacings would tend to 
 positive infinity (decoupled  Al planes) if only the vibrational
 contribution mattered.

The  opposing tendencies of the internal and vibrational contributions 
 tend to compensate; in fact, summing up 
the plain surface energy and the vibrational contribution, one arrives
at the total surface free energy depicted as a function of the
interlayer spacings in panel (c) of Fig.\ref{figure_5}. For the
Al (100) surface, the  minimum in the free energy corresponds to
 $d_{12} = -0.5$ \% and $d_{23} = -0.9$\%, 
 a compromise between the  gain of
free energy upon outward relaxation, and that
in plain surface energy upon inward relaxation.

In approaches (1) and (2) from our above list (force constants 
from zero-temperature or rescaled zero-temperature posizions), the
interlayer  spacings are  $d_{12}$=--1.5\% and $d_{23}$=--1.3 \%.
 These are rather close to the minimum of the free energy
found by
direct minimization in approach (4); indeed, with reference to
Fig.\ref{figure_4}, both approach (1) (solid line) and (2) (dashed
line)  match  rather closely the values of approach (4) (crosses),
{\it i.e.} of  the full quasi-harmonic calculation.  
Approach (3) (dash-dotted line),  where we rescaled the lattice constant and
then re-optimized all atomic positions, fails badly, going astray
already near the Debye temperature, 
and progressively more so for higher temperatures. This is due  
 to the incorrect (free-energy--wise) spacings imposed on the near-surface 
layers
by the minimization of the internal energy. The spacings are found to be 
$d_{12}$=--3.4\% and $d_{23}$=--3.1 \% at 450 K, and 
$d_{12}$=--5.8\% and $d_{23}$=--5.5 \% at 900 K.
A glance at panel (c) of Fig.\ref{figure_5} reveals that both of
these points in the \{$d_{12}, d_{23}$\} plane do indeed correspond to
free energies very far away from the minimum (especially at the higher 
temperature).

In conclusion, the most naive and simplest approach of
 exporting the $T=0$ force constants and surface energy to non-zero 
temperature does indeed underestimate  considerably the temperature
 variation of the surface free energy with respect to thermodynamic
 integration, but it gives  gives an agreement comparable to, or
 better than the sophisticated adjustment of the interlayers to find
 the free energy minimum.    

\section{Summary}
We have presented a refined Al interatomic potential for
classical dynamics and Monte Carlo simulations. We thoroughly
tested its functionalities, finding it to be very accurate for
a variety of systems. Next, we applied it to evaluating the
performance of quasi-harmonic approaches to free energy calculations
for surfaces, comparing the latter results with full thermodynamic
integration results. For Al surfaces,
the quasi-harmonic approximation shows a progressively increasing
error for temperatures above $\Theta_{D}$.
Different levels of quasi-harmonic approximation have been compared;
for Al, the simplest method of using zero-temperature phonons to
compute the free energy at all temperatures is as accurate as the 
explicit minimization of the free energy with respect to geometrical 
parameters.

\acknowledgments
We thank  Dr.\ Furio Ercolessi for helpful assistance.  
U. Hansen and P. Vogl acknowledge financial support by Siemens AG.
V. Fiorentini was supported by
the Alexander von Humboldt-Stiftung during his stay at the Walter Schottky
Institut.

\appendix
\section{Computational details}
\label{details}

\subsection{Substrate sizes}\label{substrate}
For the calculation of the surface self-diffusion barriers, and surface and
step energies, we have employed finite slabs with periodic boundary
conditions for the lateral cells. The supercells contained of 672, 550
and 560 atoms for the Al(111), Al(100) and Al(110) surfaces and
consisted of 12, 11 and 9 atomic layers. In order to determine the
surface energies of Al(111), Al(100) and Al(110) the supercells
contained 1080, 550 and 560 atoms, arranged in 9, 11 and 16 atomic
layers. The step formation energies were obtained from systems
containing 4 steps and 72, 105, 102 and  102 atoms per layer
corresponding to a total number of 1312, 2724, 1368 and 2532 atoms for
the Al$(1,0,9)$, Al$(1,1,15)$, Al$(8,8,10)$ and Al$(9,9,7)$
surface. All forces {\bf F} per atom have been brought below a
threshold of $10^{-5}$ eV/\AA. We estimated the errors in the total
energies due to the finite supercell size to be well below
$10^{-4}$ eV/atom.

\subsection{Molecular dynamics calculation of the sticking probability}
\label{stickingdetails}
The reaction probabilities were calculated in classical molecular
dynamics simulations using our Al interaction potential. The
integration was performed with a 5-th order Runge Kutta method with an
adaptive timestep, in order to ensure total energy conservation
throughout the simulation.
 Supercells containing 1320 atoms arranged in 10 atomic layers
were employed; cell dimensions are chosen so as to avoid artifacts of
the in-plane periodicity. The starting configuration is chosen to be a
(111) surface, the one Al surface with the lowest formation
energy. All atomic coordinates are allowed to evolve  dynamically,
except those of the two bottom layers of the supercell. The surface
temperature is set at 450 K (i.e about 1/2 of the melting temperature,
and $\sim$ 15 \% larger than the bulk Debye temperature). 

\subsection{Monte Carlo calculations within the canonical ensemble}
All  Monte Carlo calculations  were be performed within the canonical
ensemble, using the standard Metropolis  
technique.\cite{frenkel,allen} The
maximum atomic displacement was automatically adjusted in order to get
an acceptance ratio of 0.4. It was not systematically studied that
this acceptance ratio was an optimum, but
well converged  statistical  averages were obtained
with  a typical number of Monte
Carlo moves of order  10$^4$ times the number of atoms in the system.
 Before averaging, the system was equilibrated
for a number of steps of order
500 times the number of atoms in the system. For the Al (100) surface
we used in total 384 atoms.

\subsection{Quasi-Harmonic free energy calculations} 
Within the quasi-harmonic methods we employed slab geometries with 20
atomic layers each containing 32 atoms. For the ${\bf k}$-space
summation we used grids typically containing 2500 equally spaced ${\bf
k}$-points. Careful tests showed that this number of ${\bf k}$-points
yields well converged results.

\narrowtext
\begin{table}
\caption{Parameters used to define the atomic density function $\rho(r)$. The 
positions of the spline knots and the values at the knots are given. Also the 
first derivatives at the first and last knots are given.}\label{rh} 
\begin{tabular}{ccc}
$r$ [\AA]& $\rho(r)$ & $\rho^\prime(r)$ [1/\AA]\\
\hline
0.0000 & 0.000                 & 0.0000\\
1.8000 & 6.3820$\times10^{-1}$ & \\
1.9000 & 7.6541$\times10^{-1}$ & \\  
2.0211 & 8.6567$\times10^{-1}$ & \\
2.2737 & 9.2521$\times10^{-1}$ & \\
2.5264 & 8.6200$\times10^{-1}$ & \\
2.7790 & 7.6273$\times10^{-1}$ & \\
3.0317 & 6.0648$\times10^{-1}$ & \\
3.2843 & 4.6603$\times10^{-1}$ & \\
3.5370 & 3.3874$\times10^{-1}$ & \\
3.7896 & 2.3257$\times10^{-1}$ & \\
4.0422 & 1.0905$\times10^{-1}$ & \\
4.2949 & 5.2491$\times10^{-2}$ & \\
4.5475 & 3.9170$\times10^{-2}$ & \\
4.8001 & 3.0828$\times10^{-2}$ & \\
5.0528 & 2.5021$\times10^{-2}$ & \\
5.3054 & 1.4722$\times10^{-2}$ & \\
5.5600 & 0.0000                & 2.1298$\times10^{-6}$\\
\end{tabular}
\end{table}

\begin{table}
\caption{Parameters entering the potential energy function from 
Eq.(\protect\ref{e_tot})}\label{parameters} 
\begin{tabular}{cc}
Parameter &  \\ \hline
R$_0$ [\AA]       & 5.46    \\
D$_0$ [\AA]       & 0.10    \\
R$_\Phi$ [\AA]    & 2.00    \\
D$_\Phi$ [\AA]    & 0.25    \\
A [eV]            & 7255.44 \\
$\lambda$ [1/\AA] & 4.42085 \\
B [eV]            & 1.04897 \\
\end{tabular}
\end{table}

\begin{table}
\caption{Parameters used to define the embedding function
$F(\rho)$. The positions of the spline knots and the values at the
knots are given. Also the first derivatives at the first and last
knots are given.}\label{uu}  
\begin{tabular}{ccc}
$\rho$ & $F$($\rho$) [eV]& u$^\prime(\rho)$ [1/eV] \\ 
\hline
0.0 &    0.0000  & -12.375\\  
0.1 &   -0.8139  & \\
0.2 &   -1.2697  & \\
0.3 &   -1.6799  & \\
0.4 &   -2.0296  & \\
0.5 &   -2.2520  & \\
0.6 &   -2.4272  & \\
0.7 &   -2.5517  & \\
0.8 &   -2.6052  & \\
0.9 &   -2.6440  & \\
1.0 &   -2.6571  & \\
1.1 &   -2.6456  & \\
1.2 &   -2.6087  & \\
1.4 &   -2.4525  & 1.0620\\
\end{tabular}
\end{table}

\begin{table}
\caption{Parameters used to define the pair potential $\phi(r)$. The
positions of the spline knots and the values at the knots are
given. Also the first derivatives at the first and last knots are
given.}\label{v2}   
\begin{tabular}{ccc}
$r$ [\AA] & $\phi(r)$ [eV] & $\phi^\prime(r)$ [eV/\AA] \\ \hline
2.0211 &  1.9601    & -7.0273\\
2.2737 &  6.8272$\times10^{-1}$   & \\
2.5263 &  1.4737$\times10^{-1}$   & \\
2.7790 & -1.8818$\times10^{-2}$  & \\
3.0317 & -5.7601$\times10^{-2}$  & \\
3.2843 & -5.1984$\times10^{-2}$  & \\
3.5369 & -3.7635$\times10^{-2}$  & \\
3.7896 & -3.7373$\times10^{-2}$  & \\
4.0422 & -5.3135$\times10^{-2}$  & \\
4.2949 & -6.3286$\times10^{-2}$  & \\
4.5475 & -5.4810$\times10^{-2}$  & \\
4.8001 & -3.7288$\times10^{-2}$  & \\
5.0528 & -1.8887$\times10^{-2}$  & \\
5.3054 & -5.8523$\times10^{-3}$ & \\
5.5600 &  0.0000  & 5.9065$\times$10$^{-6}$\\
\end{tabular}
\end{table}

\narrowtext
\begin{table}[h]
\caption{Comparison of selected hopping and exchange  diffusion 
 barriers on low-index Al
surfaces obtained with the present model and in {\it ab initio}
calculations. Al (111) is also included for completeness.}
\label{diffusion_barriers} 
\begin{tabular}{lcc}
               & This work & {\it ab initio} \\ \hline
Al (111) hopping       & 0.04 & 0.04\tablenotemark[1] \\
Al (100) hopping      & 0.60 & 0.68\tablenotemark[1],0.65\tablenotemark[2]\\
Al (100) exchange    & 0.50&  0.35\tablenotemark[1] \\
Al (110) $\bot$ hopping   & 1.13 & 1.06\tablenotemark[1] \\
Al (110) $\|$ hopping & 0.30 & 0.60\tablenotemark[1] \\
\end{tabular}
\tablenotemark[1]{Reference \onlinecite{stumpf}};\,
\tablenotemark[2]{Reference \onlinecite{feibelman}} \end{table}

\narrowtext
\begin{table}
\caption{Surface formation energies for low index Al surfaces calculated with 
the present model and with other theories.}\label{surface_energies} 
\begin{tabular}{ccccc}
& \multicolumn{2}{c}{This work} & \multicolumn{2}{c}{LDA\tablenotemark[1]} \\
System   & (eV/atom) & (eV/\AA$^2$) & (eV/atom) & (eV/\AA$^2$) \\ \hline
Al(111)  & 0.38  & 0.054 & 0.48 & 0.070 \\ 
Al(100)  & 0.48  & 0.059 & 0.56 & 0.071 \\
Al(110)  & 0.74  & 0.065 & 0.89 & 0.080 \\
\end{tabular}
\tablenotemark[1]{Reference \onlinecite{stumpf2}} \end{table}
\narrowtext\begin{table}
\caption{Step formation energies for low index Al surfaces calculated with the 
present model and with other theories.}\label{step_energies} 
\begin{tabular}{ccccc}
& \multicolumn{2}{c}{This work} & \multicolumn{2}{c}{LDA\tablenotemark[1]} \\
System   & (eV/\AA) & (eV/atom) & (eV/\AA) & (eV/atom) \\ \hline
Al$[9(100) \times (111)]$ & 0.055  & 0.142 &  &  \\ 
Al$[9(100) \times (110)]$ & 0.066  & 0.240 &  &  \\
Al$[9(111) \times (111)]$ & 0.083  & 0.215 & 0.082 & 0.232 \\
Al$[9(111) \times (100)]$ & 0.085  & 0.222 <& 0.088 & 0.248 \\
\end{tabular}
\tablenotemark[1]{Reference \onlinecite{stumpf1}} \end{table}

\end{multicols}

\begin{references}
\bibitem{expansion}
The thermal expansion of the lattice constant is well described
 for 0 $< T <$  950 K by 
 $a(T) = a_0 + \alpha_1 T + \alpha3 T^3$, where $a_0$ = 4.032
\AA, $\alpha_1 = 6.35 \times 10^{-5}$ \AA/K and $\alpha_3 = 1.26
\times 10^{-11}$ \AA/K$^3$.
\end{references}
\end{document}